\documentclass[aps,preprint,prd,showpacs,nofootinbib]{revtex4}
\usepackage{amsmath}
\usepackage{graphicx}
\usepackage{dcolumn}
\usepackage{bm}
\usepackage{amssymb}
\usepackage{latexsym}

\def\be{\begin{equation}}
\def\ee{\end{equation}}
\def\ba{\begin{eqnarray}}
\def\ea{\end{eqnarray}}

\begin{document}

\title{ On Number of Nflation Fields
}

\author{Iftikhar Ahmad,
Yun-Song Piao 
and Cong-Feng Qiao
}
\affiliation{College of Physical
Sciences, Graduate University of Chinese Academy of Sciences
YuQuan Road 19A Beijing 100049, China.}

\begin{abstract}

In this paper, we study the Nflation model, in which a collection
of massive scalar fields drives the inflation simultaneously. We
observe, when the number of fields is larger than the square of
ratio of the Planck scale $M_p$ to the average value $\bar m$ of
fields masses, the slow roll inflation region will disappear. This
suggests that in order to make Nflation responsible for our
observable universe, the number of fields driving the Nflation
must be bounded by the above ratio. This result is also consistent
with recent arguments from black hole physics.

\end{abstract}

 \maketitle

\section{Introduction}

The multiple field inflation implemented by assisted inflation
mechanism proposed by Liddle et.al \cite{LMS} relaxes many limits
for the single field inflation models, and is being become a
promising class of inflation models. There has been many studies
on it \cite{MW, Piao0206}. Recently, Dimopoulos et.al \cite{DKMW}
showed that the many axion fields predicted by string vacuum can
be combined and lead to a radiatively stable inflation, called
Nflation, which may be an interesting embedding of inflation in
string theory. Then the detailed study was made by Easther and
McAllister \cite{EM} for quite specific choices of initial
conditions for the fields. In Nflation model, the spectral index
of scalar perturbation is always redder than that of its
corresponding single field, which is given numerically in
\cite{KL, SA} and is showed analytically in \cite{Piao0606}
\footnote{see also different result for Nflation with small-field
potential \cite{APQ}. }, the ratio of tensor to scalar has always
same value as in the single field case \cite{AL}, the
non-Gaussianity is quite small \cite{SA2, BB}. There was some
further studies \cite{SL}.

In single field inflation model, the occurrence of inflation
requires the value of field must be beyond the Planck scale, which
can be obtained by imposing the slow roll condition upon the
field. However, when the number of fields increases, this value
will decrease rapidly, and can be far below the Planck scale,
especially when the number of fields is quite large. This is a
remarkable and interesting point of Nflation model.

In addition, in single field inflation model, when the value of
field increases up to some value, the quantum fluctuation of field
will inevitably overwhelm its classical evolution along the
potential. In this case, the inflaton field will undergo a kind of
random walk, which will lead to the production of many new regions
with different energy densities. In some regions, the field will
wander down along its potential, so the classical variance
dominates the evolution again and then inflation is able to cease
when the field reaches its bottom. However, in another regions the
field will fluctuate up and inflation will keep on endlessly. This
so called slow roll eternal inflation \cite{V1983, L1986}, has
been studied by using the stochastic approach \cite{AAS, GLM, L90,
L92}.

The critical value of field separating the field space into the
slow roll inflation region and eternal inflation region can be
obtained by requiring the change of classical rolling of field in
unit of Hubble time equals to its quantum fluctuation. In the case
of single field, this value is far larger than the Planck scale,
and so the end value of slow roll inflation. However, in the
Nflation model, it seems that when the number of fields is added,
the total classical roll of fields is weaken, while its total
quantum fluctuation is strengthened, which will lead to this
critical value moves faster to some smaller one, which maybe bring
a bounds for the number of fields participating in inflation. Here
the `value' for multiple field means what we take is the root of
square sum of changes of all fields, because here all fields
contribute inflation, and thus the trajectory is given by the
radial motion in field space, Thus it is interesting to check this
possibility. This will be done in this paper. In section II, we
will study the case of Nflation with massive fields. Firstly we
show a simply estimate for the bound of fields number by taking
Nflation with equal mass fields as an example. Then we study a
general case with mass distribution following
Mar$\check{c}$enko-Pastur law proposed by R. Easther and L.
McAllister \cite{EM}, which further validates our result. In
section III, we discuss the case of Nflation with $\phi^4$ fields.
The summary and discussion is given in the final.

\section{Bound for $N$ of Nflation}

In Nflation model, the fields are uncoupled and potential of each
filed is $V_i={1\over 2}m_i^2\phi_i^2$. The total change of all
fields is determined by the radial motion in field space. In the
slow roll approximation, we have \ba \Delta{\phi}&= &\sqrt{
\sum_i(\Delta{\phi_i})^2} = {\sqrt{\sum_i(\dot{\phi_i})^2}\over
H}\nonumber\\ &\simeq &
M^2_{p}{\sqrt{\sum_i(m^2_i\phi_i)^2}\over\sum_im^2_i\phi^{2}_i}
,\label{e2}\ea where $\Delta{\phi_i}\simeq {|{\dot{\phi_i}}|\over
H}$ and ${\dot \phi_i}\simeq {V_i^\prime\over 3H}$ have been used,
and the factor with order one has been neglected. In the meantime,
the total quantum fluctuation of fields is \ba \delta \phi &\simeq
&\sqrt{\sum_i\left(\delta \phi_i\right)^2} \simeq \sqrt{\sum_i({H
\over 2\pi})^{2}}\nonumber\\&\simeq & {\sqrt N\over M_{p}}\sqrt
{\sum_im^2_i\phi^2_i}, \label{e15}\ea where $N$ is the number of
fields, $\delta{\phi_i}\simeq {H\over 2\pi} $ has been used and
the factor with order one has been neglected. By requiring
$\Delta\phi =\delta \phi$, we will obtain the critical point
separating the slow roll inflation region and eternal inflation
region, which is given by \ba \left(\sum_i{
m^2_i\phi^2_i}\right)^{3}
  &\simeq & {M^6_{p}\over N}\sum_i \left(m^2_i\phi_i\right)^{2}.\label{e18} \ea
In slow roll inflation region, the end of slow roll inflation
requires ${{\dot H}\over H^2}\simeq 1 $, which may be reduced to
\ba M_p^2\sum_i(m^2_i\phi_i)^2 \simeq
\left(\sum_im^2_i\phi_i^2\right)^2.\label{e22}\ea

\subsection{The case with equal masses}

\begin{figure}[t,m,u]
\begin{center}
\includegraphics{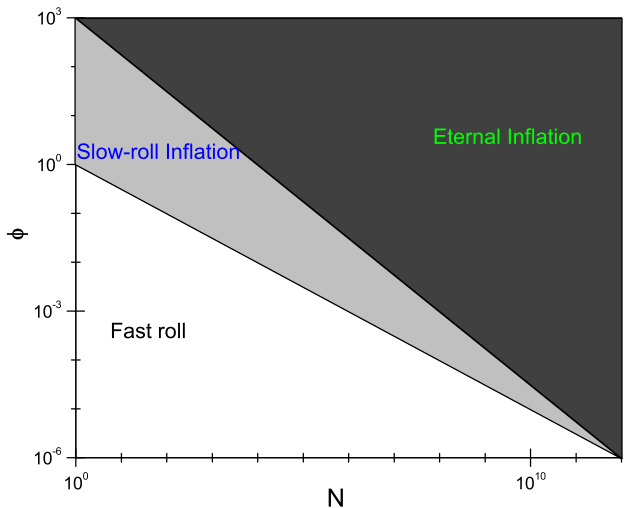}
\end{center}
\caption{ The figure of $\log$ change of the end point of slow roll
inflation and the critical point separating the slow roll inflation
region and eternal inflation region with respect to the number $N$
of fields in Nflation model with massive fields, in which $m
=10^{-6}M_p$ is taken. Three regions separated by both lines have
been pointed out in figure. We can see that when $N\sim 10^{12}$,
the slow roll inflation region disappears. }\label{xx}
\end{figure}

Firstly, when the masses of all fields are equal, i.e. $m_i=m$,
and also for simplicity we take the values of all fields are also
equal, i.e. $\phi_i=\phi$. From Eqs. (\ref{e18}) and (\ref{e22}),
we have \ba \phi &\simeq & {1\over N^{3/4}}\sqrt{M_p^3\over m},
\label{e20}\ea \ba \phi&\simeq &{M_p\over\sqrt N}, \label{e26}\ea
respectively. Thus we see that the end point moves with
${1\over\sqrt N}$, which is slower than that of the critical point
separating the slow roll inflation region and eternal inflation
region, since the latter changes with ${1\over N^{3/4}}$. This
suggests that when we plot the lines of the end point and the
critical point moving with respect to $N$, respectively, there
must be a value for these two lines to cross. Beyond this value,
the slow roll inflation region disappears as shown in Fig.
\ref{xx}. This value can be obtained by taking both Eqs.
(\ref{e20}) and (\ref{e26}) equal, which gives $N\simeq
{M_p^2\over m^2}$. Thus to make Nflation responsible for our
observable universe, the number $N$ of fields in Nflation model
must satisfy \be N\lesssim {M_p^2\over m^2}\label{e27},\ee since
the existence of such a slow roll region is significant for
solving the problems of standard cosmology and generating the
primordial perturbation seeding large scale structures of our
universe. It should be noted that in the case that the masses of
all fields are equal, their field values are equal is not reality,
because even if initially the values of all field are equal, they
will also be unequal after several efolds due to the random walk
of each field. However, this simplified analysis actually provides
a simple estimate for the bound for $N$. In next subsection, we
will validate this result in a general case.

\subsection{The case with mass distribution
following Mar$\check{c}$enko-Pastur law}

Then we will study a general case with mass distribution following
Mar$\check{c}$enko-Pastur law proposed by R. Easther and L.
McAllister \cite{EM}, which appears for axions in string theory.
The shape of the mass distribution of axions depends on the basic
structure of the mass matrix, which is specified by the
supergravity potential. In the simplest assumption, the mass
matrix is a random matrix. When one diagonalize this matrix, the
fields will be uncoupled with the mass spectrum given by the
distribution of eigenvalues. This distribution of the eigenvalues
can be characterized by Mar$\check{c}$enko-Pastur law when the
matrices are large. The mass distribution taken as the
Mar$\check{c}$enko-Pastur law is a function with respect to $\bar
m$ and $\beta$, where $\bar m$ is the average value of the mass,
i.e. $<m^2> = \bar m^2$, and $\beta$ is determined by the ratio of
the number of axions to the dimension of the moduli space and a
model dependent parameter, whose favourable value is expected to
be about $0.5$, see Refs. \cite{EM, SA}. In this case, the
smallest and largest mass are given by $m_1^2=a\equiv
\bar{m}^2(1-\sqrt\beta)^2$ and $m_{N}^2=b\equiv
\bar{m}^2(1+\sqrt\beta)^2$, respectively.

In the slow roll approximation, the field value can be given by
\be \phi_i(t)\simeq \phi_i(t_0)[\tau(t)]^{m_i^2\over b}, \ee where
$\tau(t)$ is the ratio of the value of the heaviest field at time
$t$ to its initial value $t_0$, $\tau(t)\equiv{\phi_N(t)\over
\phi_N(t_0)}$. Then defining $c\equiv {2\ln{[\tau(t)]}\over b}$,
the parts including $\phi^2_i$ in the summation terms of Eqs.
(\ref{e18}) and (\ref{e22}) can be replaced with
$\phi_i^2(t_0)\exp{[cm^2_i]}$.
When we ignore correlations between the mass distribution and the
initial field distribution, we can straightly calculate their
respective average values. By using power series expansions, the
average value of exponential term $\exp{[cm^2_i]}$ can be written
as \ba <\exp{[cm^2_i]}>=\sum_{i}<m_i^{2j}>{c^j\over
j!}.\label{ec}\ea The expectation value inside of summation in
left hand side of Eq. (\ref{ec}), can be expressed with Narayana
numbers $T(i,j)$ (see Eq. (6.14) in Ref. \cite{EM})  \ba
<m_k^{2i}> & = & \bar
m^{2i}\sum_{j=1}^{i}T(i,j)\beta^{j-1}\nonumber\\ &= & \bar
m^{2i}{_2{F}_1(1-i,-i,2,\beta)}, \label{ed}\ea 
where
$_2F_1$
is hypergeometric function.
Then Eq. (\ref{ec}) can be rewritten as
\ba<\exp[cm_k^2]>=\sum_{i=0}^{\infty}\bar m^{2i}
{_2{F}_1(1-i,-i,2,\beta)}{c^i\over i!}.\label{eg}\ea Therefore the
summations terms in Eq. (\ref{e18}) with the expectation values of
initial conditions and of distribution of mass spectrum of fields
are \ba\sum_{i}m_i^2\phi_i^2= N\alpha \bar
m^2\sum_{i=0}^{\infty}\bar m^{2i}
{_2{F}_1(-i,-i-1,2,\beta)}{c^i\over i!},\label{eh}\ea
\ba\sum_{i}m_i^4\phi_i^2= N\alpha \bar m^4\sum_{i=0}^{\infty}\bar
m^{2i}{_2{F}_1(-i-1,-i-2,2,\beta)}{c^i\over i!},\label{ei}\ea where
$\alpha\equiv<\phi_i^2(t_0)>$.

\begin{figure}[t,m,u]
\begin{center}
\includegraphics{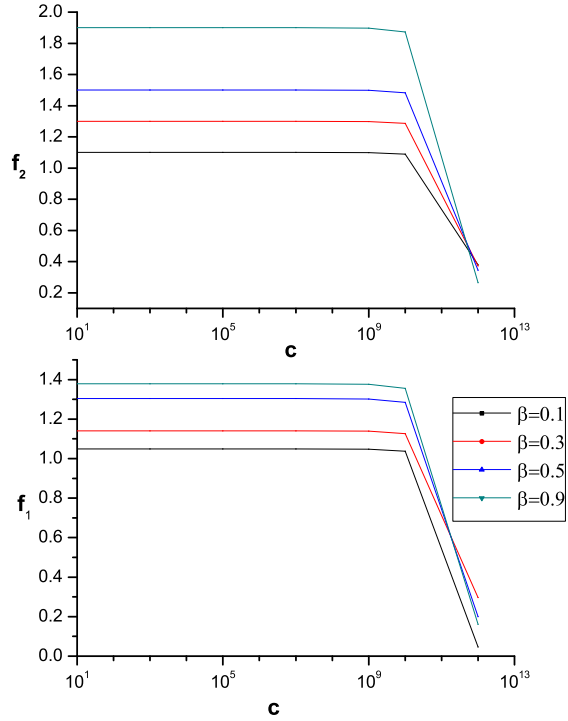}
\end{center}
\caption{ The figures of $f_1(t,\beta)$ and $f_2(t,\beta)$ with
respect to $c$ with different $\beta$. The average mass $\bar
m=10^{-6}M_p $ is taken.  }\label{yy}
\end{figure}

Thus the Eq. (\ref{e18}) with the help of Eqs. (\ref{eh}) and
(\ref{ei}) becomes\ba \alpha\simeq {M_p^3\over {\bar m} N^{3\over
2}}f_1(t,\beta),\label{ej}\ea where \ba
f_1(t,\beta)={\left[\sum_{i=0}^{\infty}\bar m^{2i}
{_2{F}_1(-i-1,-i-2,2,\beta)}{c^i\over i!}\right]^{1/2}\over
\left[\sum_{j=0}^{\infty}\bar m^{2j}
{_2{F}_1(-j,-j-1,2,\beta)}{c^j\over j!}\right]^{3/2}}, \label{ek}\ea
whose dependence on different $c$ is plotted in Fig. \ref{yy}. We
see that $f_1(t,\beta)$ is approximately a constant with order one
for a wide range of $c$, i.e. different initial conditions and
values of fields. Further, we can note that when all fields have
equal values and masses, $f_1(t,\beta)$ will have the value
$f_1(t,\beta)\simeq 1$ with $\alpha\equiv \phi^2$ and ${\bar m}=m$.
In this case Eq. (\ref{ej}) will be exactly same as Eq. (\ref{e20}).

The Eq. (\ref{e22}) with the help of Eqs. (\ref{eh}) and (\ref{ei})
 becomes \ba \alpha\simeq {M_p^2\over
N}f_2(t,\beta),\label{el}\ea where \ba
f_2(t,\beta)={\left[\sum_{i=0}^{\infty}\bar m^{2i}
{_2{F}_1(-i-1,-i-2,2,\beta)}{c^i\over i!}\right]\over
\left[\sum_{j=0}^{\infty}\bar m^{2j}
{_2{F}_1(-j,-j-1,2,\beta)}{c^j\over j!}\right]^{2}},\label{em}\ea
whose dependence on different $c$ is also plotted in Fig. \ref{yy}.
We see that similar to $f_1(t,\beta)$, $f_2(t,\beta)$ is also
approximately a constant with order one for a wide range of $c$.
When all fields have equal values and masses, $f_2(t,\beta)\simeq 1$
with $\alpha\equiv \phi^2$. In this case Eq. (\ref{el}) will be
exactly same as Eq. (\ref{e26}).

Thus combining Eqs. (\ref{ej}) and (\ref{el}) to cancel $\alpha$, we
have $N \simeq {M_p^2\over {\bar m}^2}$, where the ratio of
$f_1(t,\beta)$ to $f_2(t,\beta)$ has been taken as roughly 1, which
can be seen in Fig. \ref{yy}. This is a point in which the slow roll
inflation region will disappear. Thus to have a period of slow roll
Nflation, the number of fields must be bounded by \be N\lesssim
{M_p^2\over {\bar m}^2}. \ee This result further validates the
argument in previous subsection, only replace $m$ with $\bar m$.

\section{The case of Nflation with $\phi^4$ fields}

\begin{figure}[t]
\begin{center}
\includegraphics{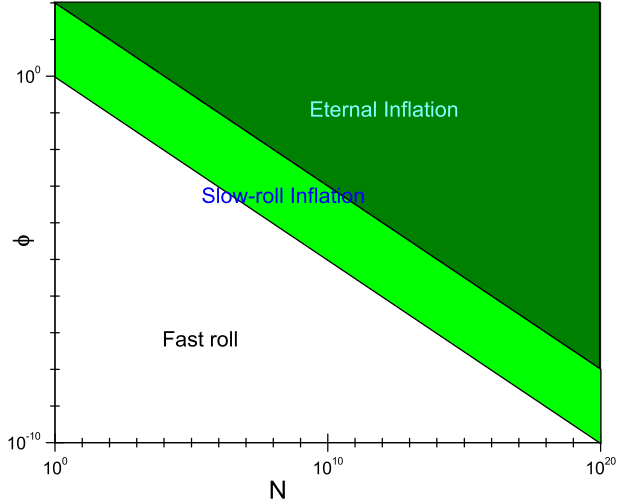}
\end{center}
\caption{The figure of $\log$ change of the end point of slow roll
inflation and the critical point separating the slow roll inflation
region and eternal inflation region with respect to the number $N$
of fields in Nflation model with $\phi^4$ fields, in which $\lambda
=10^{-12}$ is taken. Three regions separated by both lines have been
pointed out in figure. We can see that both lines are parallel, thus
there is not the bound for $N$. }\label{zz}
\end{figure}

It is interesting to further check whether there is similar bound
for the field number of Nflation with $\phi^4$ fields. Following
the same steps as we did in the previous section, the critical
point separating the slow roll inflation region and eternal
inflation region and the end point of inflation can be given by
\ba \left(\sum_i{ \lambda_i\phi^4_i}\right)^{3}\simeq {M_p^6\over
N}\sum_i(\lambda_i\phi^{3}_i)^{2}, \label{e40} \ea \ba
M_{p}^2\sum_i(\lambda_i\phi_i^{3})^2 \simeq
\left(\sum_i\lambda_i\phi_i^4\right)^2, \label{e43}\ea
respectively, where $\lambda_i$ is the couple constant of the
corresponding field and the factors with order one have been
neglected. For simplicity, we take all $\phi_i =\phi$ and
$\lambda_i = \lambda$, and thus have \be \phi \simeq  {M_p\over
\lambda^{{1\over 6}} \sqrt{N}}, \label{l1}\ee \be \phi\simeq
{M_p\over \sqrt N}. \label{l2}\ee Thus in this case the critical
point and the end point approximately obey the same evolution with
$N$ increased, which is plotted in Fig.\ref{zz}. This suggests
that there is not the bound for the number of fields imposed by
the occurrence of slow roll inflation.

This result seems unexpected. The reason leading to it may be that
for $\phi^4$ field, its effective mass is $\sim \lambda\phi^2$,
which is changed with $\phi$, and its change in some sense sets
off the fast moving of the critical point. When writing $\lambda
\phi^2 =m^2$ in Eq. (\ref{l1}), one can find that the resulting
equation will be the same as Eq. (\ref{e20}). Thus combining it
with Eq. (\ref{l2}), we will have the same result with Eq.
(\ref{e27}), which in turn suggests \be \lambda \phi^2\lesssim
{M_p^2\over N}. \label{l3}\ee Thus for $\phi^4$ field, the bound
relation between the mass $m$ and $N$ in Eq. (\ref{e27}) for
massive field is transferred to that between the field value
$\phi$ and $N$. The study with general case will be expected to
have approximately the same result with Eq. (\ref{l3}), which is
neglected here.

\section{Summary and Discussion}

In this paper, we study the Nflation model, in which a collection
of massive scalar fields drives simultaneously the inflation. We
observe that with the increase of fields number, both the critical
point separating the slow roll inflation region and eternal
inflation region and the end point of slow roll inflation will
move towards smaller average value of fields, however, at
different rates. In general, the critical point moves faster than
the end point, which leads to that the slow roll inflation region
will be eat off by the eternal inflation region inch by inch. When
the number of fields is enough large, i.e. $N\simeq M_p^2/{\bar
m}^2$, both points overlaps, which means that the slow roll
inflation region completely disappears. In this sense, in order to
make Nflation responsible for our observable universe, the field
number driving the Nflation must be bounded by $N\lesssim
M_p^2/{\bar m}^2$.


Recently, it was shown that in theories with a large number $N$ of
fields with a mass scale $m$, black hole physics imposes a bound
between $M_p$ and $N$ \cite{D07, D08}, i.e. $Nm^2\lesssim M_p^2$,
which is actually the same as the result showed here. This can be
explained as follows. In general each field can contribute the
factor $\sim m^2$ into the renormalization of the Planck mass,
thus after the accidental cancellations are neglected, the net
contribution of $N$ fields will be $\sim Nm^2\simeq M_p^2$. This
indicates that with $N$ massive fields there exists a
gravitational cutoff, beyond which the quantum gravity effect will
become important \cite{D10}. When we focus on the inflation driven
by $N$ massive fields, we observe that the same cutoff will also
appear in a similar sense, i.e. beyond this cutoff the quantum
effect will be dominated, it is which that leads to the
disappearance of the slow roll inflation region.
Thus in this sense our result also justifies the bound of Ref.
\cite{D07} from a different point of view. In this case exactly
the Plank mass should be replaced by renormalized one, i.e.
${\tilde M}_p^2$, which includes the contributions of all massive
fields for $M_p^2$. However, it can be noted that ${\tilde M}_p^2$
is actually the same order as $M_p^2$. Thus when the
renormalization of $M_p^2$ is considered, our result is not
qualitatively altered.

However, this bound can not be applied to nearly massless scalar
field. The reason is when the masses of fields are negligible,
they will not appear in summation for fields in both sides of
Eqs.(\ref{e18}) and (\ref{e22}), which is actually also a
reflection that the massless fields do not affect the motion of
massive fields dominating the evolution of universe, while the
perturbations summed in Eq.(2) are those along the trajectory of
fields space, since the massless fields only provide the entropy
perturbations orthogonal to the trajectory, which thus are not
considered. The same case can be also seen in argument of
\cite{D07}, since the contribution of field to the renormalization
of the Planck mass is proportional to $m^2$, thus the net
contribution leaded by all massless fields to the Planck mass may
actually be neglected. Thus if there are some nearly massless
fields and some massive fields with nearly same order, it should
be that there is a bound $N\lesssim M_p^2/{\bar m}^2$, in which
only massive fields are included in the definition of $\bar m$ and
$N$. In our example when the mass distribution is characterized by
Mar$\check{c}$enko-Pastur law, as has been used, in which the
smallest and largest mass are given by $m_1^2=
\bar{m}^2(1-\sqrt\beta)^2$ and $m_{N}^2=
\bar{m}^2(1+\sqrt\beta)^2$, respectively, the masses of all fields
are approximately in same order for $\beta\simeq 0.5$. In this
case it is natural that all fields need to be considered.

The field $\lambda \phi^4$ in essence is different from massive
field $m^2\phi^2$. The former corresponds to have a running mass
$\sim \lambda \phi^2$, which is dependent of $\phi$. In this case,
following Ref. \cite{D07}, the net contribution of $N$ fields to
the renormalization of the Planck mass will be $\sim N\lambda
\phi^2\simeq M_p^2$. Thus a same bound relation with Eq.(\ref{l3})
can be obtained, which is seemingly one between the field value
$\phi$ and $N$. Eq.(\ref{l3}) can be written as \be \phi \lesssim
{M_p\over \sqrt{\lambda N}}. \label{15}\ee Note that for general
$\lambda <1$, when there is a slow roll inflation region,
Eq.(\ref{15}) is always satisfied, since the inequality given by
Eq.(\ref{e40}) is actually included in Eq.(\ref{15}). Thus it
seems that there is not a bound for $N$ in the Nflation with
$\lambda\phi^4$.

In principle, the bound showed here seems be only valid for
massive scalar fields. Further, whether there are similar bounds
for other fields with various potentials remains open, and needs
to be studied. Be that as it may, however, the result observed,
that there may be a large N transition leaded by the quantum
effect in inflation, may be interesting, which might have deep
relations with other large N phenomena discussed, and thus is
worth further explore.

\textbf{Acknowledgments}

We thank S.A. Kim for a kindly help on relevant details in II.B
and also Y.F. Cai for helpful comments and discussions. I.A thanks
the support of (HEC) Pakistan. This work is supported in part by
NSFC under Grant No: 10491306, 10521003, 10775179, 10405029,
10775180, in part by the Scientific Research Fund of
GUCAS(NO.055101BM03), in part by CAS under Grant No: KJCX3-SYW-N2.

\end{document}